\documentclass[conference]{IEEEtran}

\usepackage[utf8]{inputenc}
\usepackage[T1]{fontenc}
\usepackage{graphicx}
\usepackage{amsmath,amssymb}
\usepackage{booktabs}
\usepackage{hyperref}
\usepackage{subcaption}
\usepackage{xcolor}
\usepackage{url}
\usepackage{caption}

\hypersetup{
  colorlinks=true,
  linkcolor=black,
  citecolor=black,
  urlcolor=black,
}

\newcommand{\SSIM}{\textit{SSIM}}
\newcommand{\TV}{\textit{TV}}

\begin{document}

\title{An Unsupervised Reconstruction Method For Low-Dose CT Using Deep Generative Regularization Prior}

\author{
\IEEEauthorblockN{Mehmet Ozan Unal\textsuperscript{1}, Metin Ertas\textsuperscript{2}, Isa Yildirim\textsuperscript{1}}
\\
\IEEEauthorblockA{\textsuperscript{1}Electronics and Communication Engineering Dept., Istanbul Technical University, Istanbul, Turkey\\
\textsuperscript{2}Electrical and Electronics Engineering Dept., Istanbul University, Istanbul, Turkey\\
Email: unalmehmet@itu.edu.tr, ertas@istanbul.edu.tr, iyildirim@itu.edu.tr}
}

\maketitle

\begin{abstract}
Low-dose CT imaging requires reconstruction from noisy indirect measurements which can be defined as an ill-posed linear inverse problem. In addition to conventional FBP method in CT imaging, recent compressed sensing based methods exploit handcrafted priors which are mostly simplistic and hard to determine. More recently, deep learning (DL) based methods have become popular in medical imaging field. In CT imaging, DL based methods try to learn a function that maps low-dose images to normal-dose images. Although the results of these methods are promising, their success mostly depends on the availability of high-quality massive datasets. In this study, we proposed a method that does not require any training data or a learning process. Our method exploits such an approach that deep convolutional neural networks (CNNs) generate patterns easier than the noise, therefore randomly initialized generative neural networks can be suitable priors to be used in regularizing the reconstruction. In the experiments, the proposed method is implemented with different loss function variants. Both analytical CT phantoms and human CT images are used with different views. Conventional FBP method, a popular iterative method (SART), and TV regularized SART are used in the comparisons. We demonstrated that our method with different loss function variants outperforms the other methods both qualitatively and quantitatively.
\end{abstract}


\section{Introduction}

X-Ray Computed Tomography (CT) is one of the most commonly used imaging modalities in clinical applications. CT uses ionizing radiation to monitor inside of human body. However, ionizing radiation increases the risk of radiation-related diseases such as cancer which is the biggest limitation of CT imaging. Therefore, X-ray dose reduction without sacrificing the image quality has been one of the most active research areas on this subject. Mainly, there are two proposed solutions to reduce CT radiation dose: \textit{i)} reducing the number of projections (sparse-view), \textit{ii)} reducing the x-ray tube current. In mathematical form, sparse-view CT reconstruction can be described as an ill-posed inverse problem as:

\begin{equation}
    \label{equ:yax}
    y=A x^{*}+\eta
\end{equation}

where $x^{*} \in \mathbb{R}^{n}$ is the vector form of the unknown image, $y \in \mathbb{R}^{m}$ is the measurement vector and $A \in \mathbb{R}^{m \times n}$ is the forward projection operator. Since the number of measurements ($m$) is far fewer than the number of unknowns ($n$), it is a severely underdetermined inverse problem.

In literature, Compressed Sensing (CS) based methods have been proposed to deal with underdetermined inverse problems. CS theory suggests that full recovery of the image is possible even if it is below the Nyquist barrier if such condition is met that the image should be sparse on a known basis \cite{cs-donoho,cs-candes}. To exploit this hypothesis, handcrafted regularizers are selected according to the priors of the image. The regularizers are used to constraint the solution set according to prior knowledge of the data. One of the most commonly used regularizers for natural images is Total Variation (TV) prior which assumes that the natural images are smooth and should have a smaller total gradient \cite{rof92,Sidky_tv}. Usually, these handcrafted priors are simplistic and harder to determine because of the variation of the datasets.

Iterative methods are also used to solve linear inverse problems. Simultaneous Algebraic Reconstruction Technique (SART) solves (\ref{equ:yax}) to estimate $x$ iteratively by simultaneously back projecting the error \cite{sart1984}. It is also combined with the handcrafted priors and applied with them \cite{sart_tv}.

Recently, deep learning (DL) has shown some promising results on image processing problems such as denoising, super-resolution, and inpainting \cite{dncnn-dl,superres-dl,inpainting-dl}. For sparse-view CT reconstruction, DL methods are studied with different approaches. These approaches can be categorized into three groups: \textit{i}) post-processing of sparse view image reconstructions, \textit{ii}) learning a mapping from measurement to image domain via DL, \textit{iii}) DL based iterative methods.

Post-processing of sparse view CT reconstruction method works in the image domain and approaches to this problem as a denoising problem. The first studies used supervised CNN training to solve this problem \cite{FBP+U-Net, redcnn}. In 2014, Generative Adversarial Networks (GAN) was proposed which suggests a new training approach for neural networks \cite{GoodfellowGan}. GAN has given excellent results in solving the problem of smoothness when a purely supervised learning scheme is used. In 2017, a type of GAN, Wasserstein GAN was proposed for image generation problems \cite{wasserstein-gan}. These ideas were implemented for low-dose CT reconstruction problems. GAN was applied for these problems with the combination of different losses such as perceptual loss, Wasserstein loss, pixel loss \cite{tomogan,ct-wgan-vgg}. One of the last development in GAN research is CycleGANs which also implement the backward operator to create consistency \cite{CycleGAN}. CycleGANs is also proposed to solve the CT super-resolution problem by enforcing cycle-consistency in terms of the Wasserstein distance to establish a nonlinear end-to-end mapping from noisy low-resolution input images to denoised and deblurred high-resolution outputs \cite{CycleGanCT}.

Direct reconstruction learning is such a method that learns a mapping from the measurement domain to the image domain via a deep neural network. AUTOMAP was used for this purpose in magnetic resonance imaging (MRI) reconstruction \cite{automap}. It was also pointed out that it was not feasible to use it for CT reconstruction because of the complexity of the backward operator. To overcome this complexity, He \textit{et al.} developed a method called iRadonMAP to reduce the number of required learnable parameters for a mapping from measurement to image domain to enable direct reconstruction learning applicable for CT reconstruction problem \cite{iradon}.

Deep iterative methods are another suggested solution for low-dose CT reconstruction. Earlier methods in this area suggested unrolled optimization which learns a prior during training \cite{IterativeANNforCT}. These methods try to learn an iterative scheme mostly at a certain number of iterations. Since unrolled deep neural networks are quite complex structures, it is harder to optimize and there is no guarantee for the convergence. Therefore, the studies are started to focus on solutions of this optimization problem. Alternating direction method of multipliers (ADMM) breaks complex convex optimization problems into smaller pieces to deal with complex structures \cite{ADMM}. ADMM method was also used to learn regularizer in CT image reconstruction \cite{ADMMNET}. Adler and Oktem suggested an optimization method, learned primal-dual, \cite{adler-2017-lpdr} to solve this iterative scheme which claims that it is data-efficient and time-effective.

All three approaches mentioned above usually require the availability of clean, massive, labeled datasets. For medical imaging, it is quite challenging to collect clean data to use as training labels. Evaluation of the datasets requires tedious work by domain experts, in our case radiologists. DL based methods have also been exploited in an unsupervised sense as regularizers. Bora \textit{et al.} suggested using generative models as reconstruction priors \cite{bora2017}. Deep image prior (DIP) exploited the method of deep generative network regularization for several image processing problems such as super-resolution, denoising, inpainting \cite{Ulyanov_2018_CVPR}. DIP method has been applied on various fields: positron emission tomography (PET) reconstruction \cite{pet-dip}, diffraction tomography \cite{difftomo-dip} and image restoration \cite{ImageRestoration-tv-dip}. Recently, DIP was tested for low-dose CT reconstruction problem and compared with data-driven methods \cite{bremen-dip}. However, in that study DIP was used as an image domain denoiser.

In this paper, we developed an unsupervised method for low-dose CT reconstruction based on deep generative regularizers (DGR) which do not require any learning process or labeled big dataset. Our method theoretically relies on such a fact that deep CNNs generate patterns earlier than the random noise\cite{Ulyanov_2018_CVPR,bora2017}. Therefore, to regularize the reconstructed image in low-dose CT imaging, randomly initialized generative neural networks can be useful priors.

We investigated the potential of our proposed method for sparse CT reconstruction problem by defining a hybrid loss function that combines the losses from both the measurement domain and the image domain. The effect of each loss term is evaluated with different projection numbers on both analytical phantoms and human CT images.

\color{black}
This paper is structured as follows. In section 2, the description and realization of DGR method are given. In section 3, experiment datasets and settings and the results are given. The effect of noise level and network architecture are discussed in section 4. Finally, the paper is concluded in the last section.
\color{black}

\section{Method}

\begin{figure*}
    \includegraphics[width=0.9\textwidth]{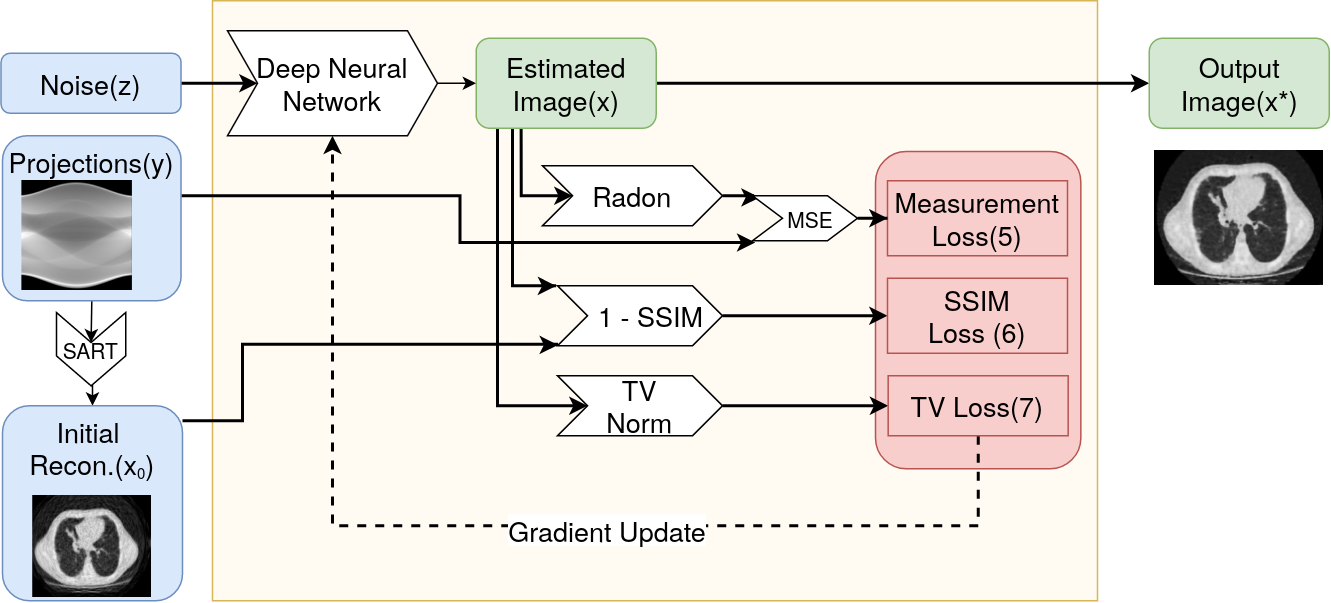}
    \caption{The proposed DGR based sparse-view CT reconstruction method working schema.}
    \label{fig:radon_dip_schema}
\end{figure*}

The generic form of the suggested optimization scheme to solve sparse view CT imaging problem can be defined as:

\begin{equation}
    \label{equ:generic}
    x^{*}= \arg \min _{x} E( Ax; y)+R(x),
\end{equation}


where $E( Ax; y)$ is the data fidelity term that penalizes the $\ell_2$ distance from the measurements.\color{black} $A$ is forward transform matrix whose coefficients are derived via Radon transform \cite{radon20051}.\color{black} The regularizer $R(x)$ is selected to constraint the solution set based on prior information. For CT imaging, to solve (\ref{equ:yax}) without a regularizer, traditionally FBP has been used to find an estimate of $x$ from the measurements, $y$. Later, regularized solutions were developed by solving (\ref{equ:generic}) in iterative manners. SART is one of the most popular iterative methods, due to its parallel programmable structure. As the regularizer ($R(x)$), TV prior is commonly used which constraints that the gradient magnitude (TV norm) of a natural image should be small. More recently, DL based approaches have been proposed to deal with sparse view CT imaging and they reported promising results. However, these methods are supervised and their success heavily depends on the availability of a clean, big, labeled dataset. An alternative way to use deep neural networks for reconstruction problems is deep generative regularizers (DGR). Deep CNNs recover patterns easier than random signals, therefore randomly initialized generative neural networks can be suitable priors to be used in regularizing the reconstruction. In this study, we aim to design such a solution that exploits this fact and does not rely on training data by proposing an unsupervised method for sparse view CT imaging which can be formulated as follows:

\begin{equation}
    \begin{split}
    \label{equ:full-proposed}
    \theta^{*} = \arg \min _{\theta} \Big\{ \ E( A G_{\theta}(z) ; y ) + \ 1-\SSIM (G_{\theta}(z),x_0) \\ + \ \TV(G_{\theta}(z)) \ \Big\}, \quad x^{*}=G_{\theta^{*}}(z) 
    \end{split}
\end{equation}

where $E$ is error function, $G$ is a deep generative neural network which is initialized with random parameters $\theta$, $z$ is the input of the generator which is randomly initialized noise, $A$ is Radon transform which is defined according to the geometry of the system, $y$ is measurements, $\SSIM$ is structural similarity function, $x_0$ is a noisy estimation of real $x$. The working schema of the proposed method is shown in Fig. \ref{fig:radon_dip_schema}. The inputs of the system are randomly initialized Gaussian noise ($z \sim \mathcal{N}(0,1)$), projections $y$ and initial reconstruction, $x_o$ which is reconstructed by SART. The output of the system is $x^*$ which is obtained by minimizing the loss function given in (\ref{equ:full-proposed}) via updating the parameters of $G$. We created a hybrid loss term that contains three different parts that constraint the solution set in different spaces. The first term of the loss function penalizes the difference from the measurements directly on the sinogram (measurement) domain. Measurements are the only raw source of data. Therefore a pixel loss is selected to penalize any inconsistency with the measurements. The second term uses an $\SSIM$ loss to constraint the solution set by the initial reconstruction. Penalizing with structural similarity helps the generator to create an image that has similar structures with the initial reconstruction. The third term penalizes the gradient magnitude of the image, which is also known as TV prior and it is one of the strongest prior for medical images. \color{black} In brief, the steps below can be followed to implement the proposed algorithm: 
\begin{itemize}
    \item The inputs of the reconstruction are the projections ($y$) and the initial reconstruction $x_0$ is calculated from $y$ via SART method. The forward operator $A$ can be derived from the system geometry.
    \item Generator neural network ($G$) is initialized randomly and the input of the network ($z$) is generated randomly from Gaussian noise.
    \item The estimated image is iteratively calculated using the loss function which is given in (\ref{equ:full-proposed})
    \item The loss function is optimized via Adam\cite{Adam}, using only these projections and the initial reconstruction.
    \item In the final step, the output can be generated using $G_{\theta^*}(z)$.
\end{itemize}

The proposed hybrid loss function in (\ref{equ:full-proposed}) will be scrutinized term by term:

\color{black}

\begin{equation}
    \label{equ:meas_loss}
    \ell_{meas} = E( A G_{\theta}(z) ; y )
\end{equation}

In (\ref{equ:meas_loss}), the measurement loss is given. $G$ is the generative neural network that is parameterized with $\theta$, $z$ is randomly initialized input of the system, $A$ is Radon transform which is defined according to the geometry of the system and $y$ is the measurements. 
\color{black} $G_{\theta}(z)$, which is the generated image is projected to sinogram domain via $A$ operator. In our geometry settings, $A$ is the forward projection matrix whose coefficients are derived via Radon transform.
\color{black}

As a generative network, two different networks are tested as U-Net \cite{Unet}, and SkipNet \cite{SkipNet}. The image ($G_{\theta}(z)$) is generated by $G$ and it is projected onto the measurement domain via Radon transform, $AG_{\theta}(z)$. Mean Squared Error (MSE) is used to measure the distance of the measurements(y) with $A G_{\theta}(z)$.

The second term of the loss function is a $\SSIM$ loss. $\SSIM$ is proposed by \cite{metric-ssim} as a metric to calculate structural similarity between two images. \color{black}The MSE metric which calculates the pixel-by-pixel differences is not necessarily correlated with perceptual similarity of the images. In order to deal with this problem, the SSIM metric has been proposed which actually aims to measure the combination of luminance, contrast and structure similarities of the images. In other words, $\SSIM$ focuses on perceptual differences between two images rather than pixel level differences. $\SSIM$ can be calculated as follows:

\begin{equation}
    \SSIM(x,y) = \frac{(2\mu_x\mu_y + C_1) + (2 \sigma _{xy} + C_2)} 
      {(\mu_x^2 + \mu_y^2+C_1) (\sigma_x^2 + \sigma_y^2+C_2)}
    \label{eq:SSMI}
\end{equation}

where $x$, $y$ are the two images whose similarity is calculated, $\mu_x$, $\mu_y$ are the averages of the images, $\sigma_x^2$, $\sigma_y^2$ are the variances of the images, $\sigma _{xy}$ is cross covariance of the images and, $ C_1$,  $C_2$ are two variables which stabilize the division. \color{black} The second term of the loss function is:

\begin{equation}
    \label{equ:ssim_loss}
    \ell_{ssim} = 1 - \SSIM(G_{\theta}(z),x_0)
\end{equation}

\setlength{\tabcolsep}{3.5pt}
\begin{table*}[t]
    \caption{ Ellipses Dataset results of changing the weights of proposed loss function terms (\ref{equ:lossfull}). SSIM values are multiplied by 100.
    }
    \footnotesize
    \begin{center}
    \begin{tabular}{lcccccccccc}
    \hline\hline
                     & \multicolumn{1}{l}{}     & \multicolumn{1}{l}{}     & \multicolumn{1}{l}{}   & \multicolumn{2}{c}{32 views}                        & \multicolumn{2}{c}{64 views}                        & \multicolumn{2}{c}{100 views}                       \\
                     & \multicolumn{1}{l}{$w_{meas}$} & \multicolumn{1}{l}{$w_{ssim}$} & \multicolumn{1}{l}{$w_{TV}$} & \multicolumn{1}{c}{PSNR} & \multicolumn{1}{c}{SSIM} & \multicolumn{1}{c}{PSNR} & \multicolumn{1}{c}{SSIM(\%)} & \multicolumn{1}{c}{PSNR} & \multicolumn{1}{c}{SSIM(\%)} \\
    \hline
    FBP              & \multicolumn{1}{l}{}     & \multicolumn{1}{l}{}     & \multicolumn{1}{l}{}   & $17.16\pm1.57$               & $33.91\pm2.28$               & $20.85\pm1.54$               & $45.56\pm3.92$               & $23.16\pm1.56$               & $55.28\pm5.07$               \\
    SART             & \multicolumn{1}{l}{}     & \multicolumn{1}{l}{}     & \multicolumn{1}{l}{}   & $24.82\pm1.96$               & $70.11\pm5.94$               & $25.76\pm1.72$               & $71.19\pm5.42$               & $25.59\pm1.65$               & $69.78\pm5.60$               \\
    SART+TV          & \multicolumn{1}{l}{}     & \multicolumn{1}{l}{}     & \multicolumn{1}{l}{}   & $26.19\pm2.15$               & $82.10\pm4.52$               & $28.03\pm1.97$               & $88.23\pm2.38$               & $28.71\pm1.86$               & $89.63\pm1.97$               \\
    DGR              & 1.000                    & 0.000                    & 0.000                  & $26.08\pm1.98$               & $81.62\pm5.77$               & $28.24\pm1.77$               & $89.49\pm2.47$               & $28.74\pm1.74$               & $91.60\pm1.51$               \\
    DGR              & 0.999                    & 0.001                    & 0.000                  & $26.66\pm1.96$               & $84.21\pm4.94$               & $28.30\pm1.78$               & $89.76\pm2.79$               & $28.75\pm1.78$               & $91.65\pm1.53$               \\
    DGR              & 0.990                    & 0.010                    & 0.000                  & $26.46\pm1.94$               & $84.02\pm3.61$               & $28.31\pm1.94$               & $88.77\pm3.10$               & $28.98\pm1.88$               & $90.46\pm2.70$               \\
    DGR              & 0.990                    & 0.000                    & 0.010                  & $26.36\pm1.87$               & $83.08\pm4.78$               & $28.36\pm1.80$               & $89.84\pm2.38$               & $28.72\pm1.74$               & $91.35\pm1.58$               \\
    DGR              & 0.980                    & 0.010                    & 0.010                  & $26.47\pm1.93$               & $83.98\pm3.46$               & $28.28\pm1.94$               & $88.65\pm3.11$               & $\mathbf{29.02\pm1.92}$      & $90.56\pm2.72$               \\
    DGR              & 0.900                    & 0.100                    & 0.000                  & $26.02\pm2.08$               & $79.42\pm4.60$               & $27.82\pm1.99$               & $82.40\pm4.62$               & $28.56\pm1.82$               & $83.93\pm4.69$               \\
    DGR              & 0.900                    & 0.000                    & 0.100                  & $\mathbf{26.82\pm1.59}$      & $\mathbf{85.43\pm2.75}$      & $\mathbf{28.45\pm1.70}$      & $\mathbf{90.30\pm1.84}$      & $28.87\pm1.68$               & $\mathbf{91.90\pm1.28}$      \\
    DGR              & 0.800                    & 0.100                    & 0.100                  & $26.02\pm2.06$               & $79.16\pm4.84$               & $27.96\pm1.97$               & $82.60\pm4.58$               & $28.64\pm1.78$               & $83.83\pm4.57$               \\
    DGR              & 0.500                    & 0.500                    & 0.000                  & $24.71\pm2.03$               & $67.84\pm6.46$               & $26.14\pm1.75$               & $68.78\pm5.43$               & $26.43\pm1.58$               & $68.30\pm4.41$               \\
    DGR              & 0.330                    & 0.330                    & 0.330                  & $24.77\pm2.02$               & $68.25\pm6.41$               & $26.20\pm1.76$               & $69.01\pm5.25$               & $26.54\pm1.56$               & $68.49\pm4.28$               \\
    DGR              & 0.000                    & 1.000                    & 0.000                  & $24.40\pm2.05$               & $66.26\pm6.48$               & $25.79\pm1.75$               & $66.41\pm5.45$               & $25.99\pm1.59$               & $65.35\pm4.91$               \\
    DGR              & 0.000                    & 0.990                    & 0.010                  & $24.43\pm2.08$               & $65.92\pm6.74$               & $25.81\pm1.75$               & $66.54\pm5.62$               & $26.01\pm1.61$               & $65.46\pm4.87$               \\
    DGR              & 0.000                    & 0.900                    & 0.100                  & $24.40\pm2.10$               & $65.89\pm7.23$               & $25.83\pm1.78$               & $66.55\pm5.67$               & $26.02\pm1.57$               & $65.73\pm4.81$               \\
    \hline\hline
    \end{tabular}
    \end{center}
    \label{table:loss_term_ellipses}
\end{table*}

\setlength{\tabcolsep}{3.5pt}
\begin{table*}[t]
    \caption{ Deep Lesion dataset results of changing the weights of proposed loss function terms (\ref{equ:lossfull}). SSIM values are multiplied by 100.
    }
    \footnotesize
    \begin{center}
    \begin{tabular}{lcccccccccc}
    \hline\hline
                     & \multicolumn{1}{l}{}     & \multicolumn{1}{l}{}     & \multicolumn{1}{l}{}   & \multicolumn{2}{c}{32 views}                        & \multicolumn{2}{c}{64 views}                        & \multicolumn{2}{c}{100 views}                       \\
                     & \multicolumn{1}{l}{$w_{meas}$} & \multicolumn{1}{l}{$w_{ssim}$} & \multicolumn{1}{l}{$w_{TV}$} & \multicolumn{1}{c}{PSNR} & \multicolumn{1}{c}{SSIM(\%)} & \multicolumn{1}{c}{PSNR} & \multicolumn{1}{c}{SSIM(\%)} & \multicolumn{1}{c}{PSNR} & \multicolumn{1}{c}{SSIM(\%)} \\
    \hline                     
    FBP              & \multicolumn{1}{l}{}     & \multicolumn{1}{l}{}     & \multicolumn{1}{l}{}   & $17.44\pm0.58$               & $35.41\pm1.52$               & $21.84\pm0.57$               & $49.82\pm1.90$               & $24.86\pm0.61$               & $63.38\pm2.13$               \\
    SART             & \multicolumn{1}{l}{}     & \multicolumn{1}{l}{}     & \multicolumn{1}{l}{}   & $25.16\pm0.94$               & $71.28\pm4.19$               & $27.46\pm0.80$               & $78.55\pm2.61$               & $28.54\pm0.80$               & $80.83\pm2.59$               \\
    SART+TV          & \multicolumn{1}{l}{}     & \multicolumn{1}{l}{}     & \multicolumn{1}{l}{}   & $24.58\pm0.84$               & $80.92\pm2.13$               & $26.05\pm0.82$               & $86.72\pm1.19$               & $26.68\pm0.83$               & $87.95\pm1.25$               \\
    DGR              & 1.000                    & 0.000                    & 0.000                  & $24.77\pm0.78$               & $85.15\pm2.19$               & $26.08\pm0.91$               & $90.07\pm1.46$               & $26.25\pm0.92$               & $91.22\pm1.16$               \\
    DGR              & 0.999                    & 0.001                    & 0.000                  & $25.20\pm0.84$               & $86.78\pm1.95$               & $26.15\pm0.91$               & $90.45\pm1.10$               & $26.29\pm0.94$               & $90.85\pm1.64$               \\
    DGR              & 0.990                    & 0.010                    & 0.000                  & $25.84\pm0.81$               & $88.09\pm1.34$               & $26.83\pm0.98$               & $\mathbf{91.64\pm0.87}$      & $27.13\pm1.04$               & $\mathbf{92.61\pm0.74}$      \\
    DGR              & 0.990                    & 0.000                    & 0.010                  & $25.06\pm0.76$               & $86.13\pm1.87$               & $26.20\pm0.93$               & $89.85\pm1.11$               & $26.36\pm0.97$               & $90.98\pm1.65$               \\
    DGR              & 0.980                    & 0.010                    & 0.010                  & $\mathbf{25.85\pm0.89}$      & $\mathbf{88.11\pm1.46}$      & $26.91\pm1.01$               & $91.47\pm0.98$               & $27.14\pm1.03$               & $92.58\pm1.12$               \\
    DGR              & 0.900                    & 0.100                    & 0.000                  & $25.66\pm0.95$               & $80.90\pm2.38$               & $28.02\pm0.98$               & $88.59\pm2.23$               & $28.99\pm1.06$               & $91.86\pm1.66$               \\
    DGR              & 0.900                    & 0.000                    & 0.100                  & $25.51\pm0.89$               & $87.02\pm1.99$               & $26.45\pm0.90$               & $90.55\pm1.20$               & $26.61\pm1.02$               & $91.36\pm1.28$               \\
    DGR              & 0.800                    & 0.100                    & 0.100                  & $25.66\pm0.94$               & $80.71\pm2.10$               & $\mathbf{28.10\pm1.04}$      & $87.94\pm2.42$               & $\mathbf{29.38\pm1.06}$      & $91.88\pm1.71$               \\
    DGR              & 0.500                    & 0.500                    & 0.000                  & $24.31\pm0.99$               & $60.70\pm4.18$               & $26.68\pm0.85$               & $66.10\pm3.12$               & $27.93\pm0.70$               & $68.96\pm1.74$               \\
    DGR              & 0.330                    & 0.330                    & 0.330                  & $24.30\pm0.97$               & $60.43\pm4.12$               & $26.74\pm0.86$               & $65.88\pm2.97$               & $28.16\pm0.77$               & $69.14\pm1.92$               \\
    DGR              & 0.000                    & 1.000                    & 0.000                  & $23.82\pm1.05$               & $54.67\pm5.58$               & $26.25\pm0.83$               & $60.12\pm4.46$               & $27.57\pm0.60$               & $63.07\pm2.62$               \\
    DGR              & 0.000                    & 0.990                    & 0.010                  & $23.81\pm1.03$               & $55.07\pm5.84$               & $26.22\pm0.80$               & $60.17\pm4.59$               & $27.50\pm0.63$               & $62.96\pm2.98$               \\
    DGR              & 0.000                    & 0.900                    & 0.100                  & $23.85\pm1.05$               & $55.42\pm5.32$               & $26.23\pm0.80$               & $60.22\pm4.50$               & $27.53\pm0.64$               & $63.12\pm2.88$               \\
    \hline\hline
    \end{tabular}
    \end{center}
    \label{table:loss_term_human}
\end{table*}

where $\SSIM$ is structural similarity, $G_{\theta}(z)$ is generated image by the neural network, $x_0$ is the initial estimation of the image by a conventional method (in our case SART method \cite{sart1984}). \color{black} $\SSIM$ metric is ranged between $0-1$, where the similarity increases as the value gets closer to $1$. Since the loss function is to be minimized, structural dissimilarity should be measured between the images. Therefore the $\SSIM$ loss is formulated as in (\ref{equ:ssim_loss}).\color{black}  

The third term of the loss function is TV norm of the image (\ref{equ:tv_loss}). This term is used to restrict the gradient magnitude of the image which constrains the piecewise smoothness.
\color{black}
TV norm of an image can be defined as follows:
\begin{equation}
    TV(x) = \sum_{i,j} \sqrt{|x_{i+1,j} - x_{i,j}|^2 + |x_{i,j+1} - x_{i,j}|^2}
\end{equation}

where $x$ is the image and $i$, $j$ are the coordinates of the image in the axial plane. 
\color{black}

\begin{equation}
    \label{equ:tv_loss}
    \ell_{TV} = \TV( G_{\theta}(z)  )
\end{equation}

The terms of the loss function are weighted differently to evaluate their contribution to the reconstruction. The final form of the loss function is as follows:

\begin{equation}
    \label{equ:lossfull}
    \begin{split}
        \ell_{total} = w_{meas} * \ell_{meas} + w_{ssim} * \ell_{ssim} + w_{TV} * \ell_{TV}, \\
        \quad w_{meas} + w_{ssim} + w_{TV} = 1
    \end{split}
\end{equation}

where $w_{meas}$, $w_{ssim}$, $w_{TV}$ are the weights of the loss function terms. The different combinations of the weights might help to enhance reconstruction performance according to different priorities.

\section{Experiments}

The source code and the experiments are available at here\footnote{https://github.com/mozanunal/SparseCT}. In these experiments, PyTorch \cite{pytorch}, a deep learning library is used with Python programming language. Scikit-image library is used for the implementation of FBP, SART and TV methods \cite{scikit-image}.

\subsection{Experiment Datasets}

The projections are generated from 2D CT slices by Radon transform. After Radon transform, zero-mean Gaussian noise is added to create noisy projections. During the experiments, uniformly distributed 32, 64, and 100 projections between $0-\pi$ are generated. The image resolution is selected as $512 \times 512$.

The experiments are done with both analytical and human CT image datasets. As analytical phantoms, Shepp-Logan and ellipses datasets are used. The images of the ellipses dataset are created with ellipses whose positions, sizes, shapes, and intensities are randomly generated. As human CT data, deep lesion dataset is used \cite{yan2017deeplesion}. In this dataset, the images are stored as 2D CT slices and the intensity values are stored as $16$ bit Hounsfield Unit (HU) values. The intensity values are normalized between $[0,1]$ for the experiments.

\subsection{Experiment Settings}

SkipNet \cite{SkipNet} is selected as denoiser neural network architecture. In the experiment architecture, the number of channels along scales are $[32, 64, 128, 256, 512]$ during downscaling and $[512, 256, 128, 64, 32]$ during the upscaling. The total number of parameters is around two million in this setting and there are a sufficient amount of parameters to overfit a single image. The parameters of the network are randomly initialized at the beginning of the reconstruction.

For the proposed DGR reconstruction, the number of iteration is selected as $4000$ and Adam optimizer \cite{Adam} is used. At the beginning of the iterations, a random Gaussian noise is created and used as the initial input of the network. During the optimization process, the initial input noise is perturbed with a zero-mean Gaussian noise whose power is significantly lower than the initial input noise. By this method, the input of the network is slightly modified at every iteration. The variance of the perturbation noise is a hyperparameter and affects the performance of the method. During the experiments, it is selected as $0.01$.

Our method is compared with FBP \cite{fbp-computed-tomography}, SART \cite{sart1984}, and SART+TV \cite{rof92} methods. For the experiments, the hyperparameters of these methods are determined empirically via selecting the values which maximize SSIM metric. For SART, the iteration number is selected as $40$ and the relaxation coefficient is selected as $0.15$. For SART+TV, TV weight is selected as $0.9$.

\begin{figure*}
    \includegraphics[width=0.98\textwidth]{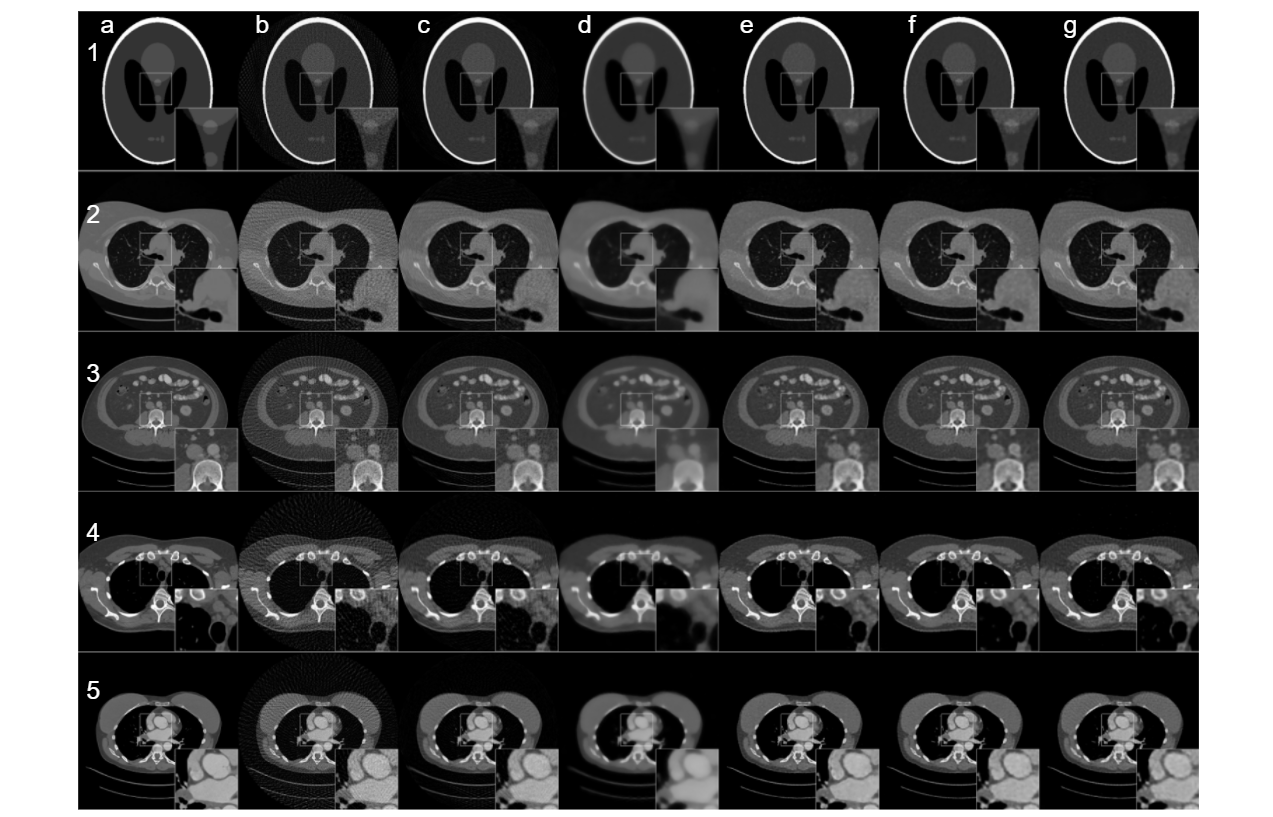}
    \caption{The images are from left to right: ground truth, FBP, SART, SART+TV, DGR ($w_{meas}$=1.0, $w_{ssim}$=0.0, $w_{TV}$=0.0), DGR ($w_{meas}$=0.9, $w_{ssim}$=0.0, $w_{TV}$=0.1), DGR ($w_{meas}$=0.98, $w_{ssim}$=0.01, $w_{TV}$=0.01). These reconstructions are obtained from 64 views with 39 dB AWGN noise.    }
    \label{fig:all-human}
\end{figure*}

\subsection{Quantitative Results}

One of the aims of the study is to determine the effect of different loss terms. To mitigate the effect of the loss function terms, a series of experiments are done. In these experiments, the loss terms are weighted at different ratios. For benchmarks, $10$ images from the ellipses dataset are generated and $10$ images are selected from deep lesion dataset. The peak signal to noise ratio (PSNR) and Structural Similarity (SSIM) \cite{metric-ssim} metrics are calculated for the experiments and they are given in  Table \ref{table:loss_term_ellipses} (for ellipses benchmark) and Table \ref{table:loss_term_human} (for deep lesion benchmark ). The loss weights are tested with 13 different combinations. For both datasets, the proposed DGR method outperforms FBP, SART, and SART+TV by generating higher PSNR and SSIM values. Since measurement loss is used in projection domain where the direct information about the object exists in, we foresee to prioritize its weight which is supported by the results given in Tables \ref{table:loss_term_ellipses} and \ref{table:loss_term_human}. From the experiments, it can be seen that combining measurement loss with the image domain and TV losses enhances the performance of the method. DGR gives the best results with weights ($w_{meas}=0.9$, $w_{ssim}=0.0$, $w_{TV}=0.1$) for ellipses dataset and with weights ($w_{meas}=0.98$, $w_{ssim}=0.01$, $w_{TV}=0.01$) for deep lesion dataset.


\begin{figure*}
    \includegraphics[width=0.98\textwidth]{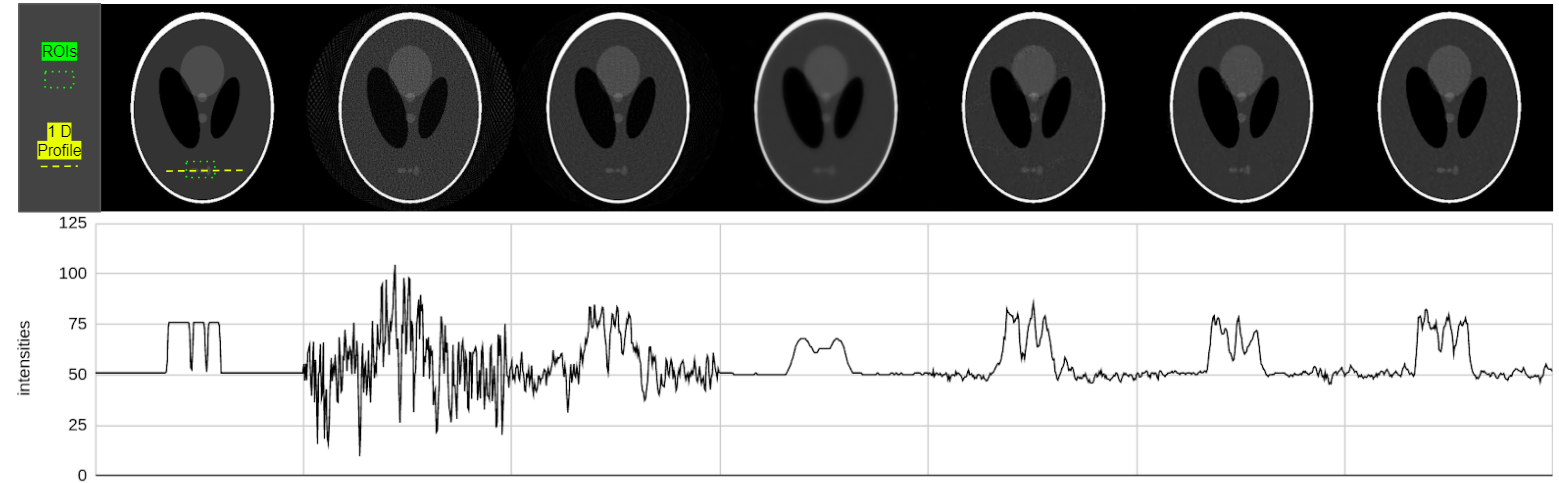}
    \includegraphics[width=0.98\textwidth]{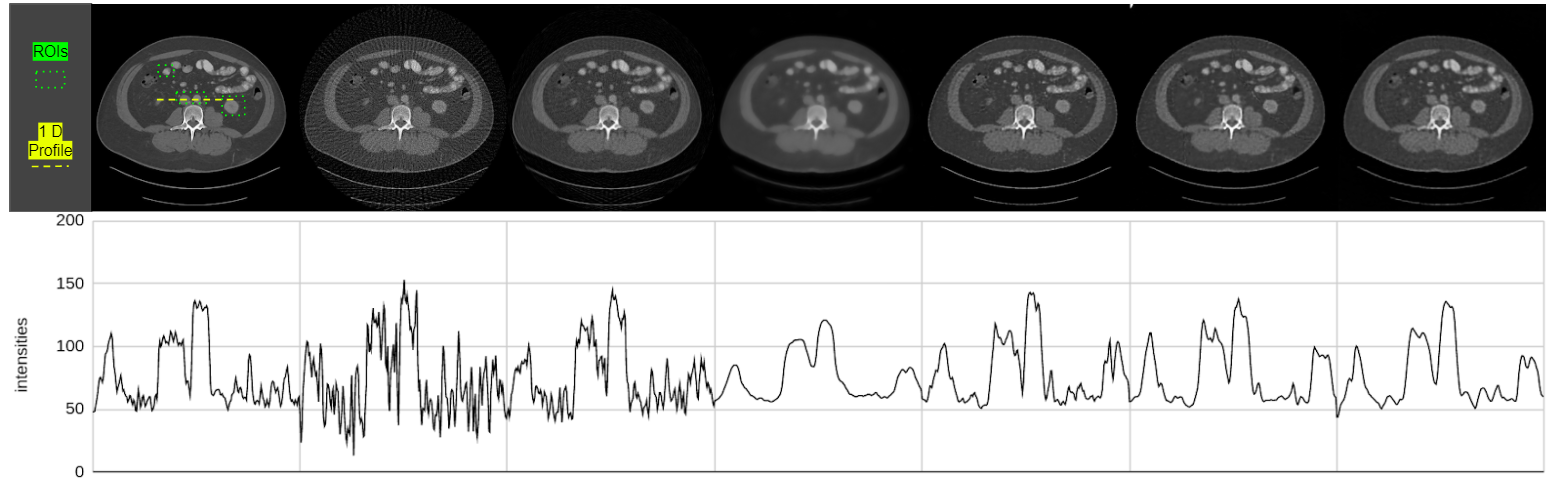}
    \caption{ 1D profiles of the reconstruction results.The images are from left to right: ground truth, FBP, SART, SART+TV, $DGR_{l_1}$ ($w_{meas}$=1.0, $w_{ssim}$=0.0, $w_{TV}$=0.0), $DGR_{l_2}$ ($w_{meas}$=0.9, $w_{ssim}$=0.0, $w_{TV}$=0.1), $DGR_{l_3}$ ($w_{meas}$=0.98, $w_{ssim}$=0.01, $w_{TV}$=0.01). These reconstructions are obtained from 64 views with 39 dB AWGN noise. \color{black} (Please see the original manuscript for the colorful marker details.)\color{black}}
    \label{fig:1d}
\end{figure*}


\subsection{Visual Results}

The visual results of Shepp-Logan phantom and four human CT images are given in Fig.\ref{fig:all-human}. Various human CT images are used in the comparisons to evaluate the performance of the methods on different tissue structures such as soft and hard tissues, lesions and tiny details. These reconstructions are obtained from $64$ views with $39$ dB AWGN (Additive White Gaussian Noise). Three different DGR variations are selected for these experiments: ($w_{meas}=1.0$, $w_{ssim}=0.0$, $w_{TV}=0.0$), DGR with measurement and TV losses combined ($w_{meas}=0.9$, $w_{ssim}=0.0$, $w_{TV}=0.1)$ and DGR with measurement, SSIM, and TV losses combined ($w_{meas}=0.98$, $w_{ssim}=0.01$, $w_{TV}=0.01)$. FBP and SART generate higher background noise. While TV is successful in suppressing background noise arising from SART, it suffers from oversmoothing fine details. However, DGR method is better than the other methods in preserving fine details and suppressing the background noise trade-off.Some parts of the reconstructed images are zoomed in Fig.\ref{fig:all-human} to take a closer look at the reconstruction performances. In row 5, in the zoomed-in image the fine details are accurately recovered via the proposed DGR methods specifically in 5-e. Spherical feature is also well separated from the features surrounding it with sharper boundaries than state-of-the-art SART+TV method in 5-d. Though SART in 5-c is successful to some extend in this sense, it suffers from very high background noise. A similar analogy is observed along the other fine details in the test data from rows 1 to 4.

In Fig. \ref{fig:1d}, 1D profiles and regions of interest for CNR calculations of Shepp-Logan phantom and human CT reconstructions of Fig. \ref{fig:all-human}.1 and \ref{fig:all-human}.5 are shown. As can be seen from Shepp-Logan 1D profiles, DGR methods are the only ones that can recover the bottom three features separately diagnosable. Compared to SART+TV, background noise is slightly higher in DGR reconstructions. For human CT data, 1D profiles of DGR methods are favorable than those of FBP, SART, and SART+TV in suppressing the noise and keeping the edges sharper trade-off. CNR results are given in Table \ref{table:CNR} and calculated with the following equation:

\begin{equation}
    \label{equ:cnr}
    CNR = \dfrac{Contrast}{Noise} = \dfrac{|\mu_{feature} - \mu_{background}|}{ \sigma_{background} },
\end{equation}

where $\mu_{feature}$ and $\mu_{background}$ are the mean intensity values of the feature and background regions, $\sigma_{background}$ is the standard deviation of the background. For both Shepp-Logan and human CT reconstructions, DGR methods outperform both FBP and SART. However, their CNR performances are slightly worse than that of SART+TV due to TV's oversmoothing effect on background.

\setlength{\tabcolsep}{4.5 pt}
\begin{table}[]
    \caption{CNR results of ROIs shown in fig \ref{fig:1d} }
    \small
    \begin{center}
        \begin{tabular}{ccccccccc}
            \hline\hline \\
            \multicolumn{1}{l}{} & FBP   & SART & TV & $DGR_{l_1}$ & $DGR_{l_2}$ & $DGR_{l_3}$ \\
            \hline
            SheppLogan          & 5.8   & 12.8 & 26.5 & 23.4 & 20.4 & 22.1 \\
            CT image             & 9.7   & 18.6 & 28.3 & 24.7 & 25.1 & 26.5 \\
            \hline\hline
        \end{tabular}
    \end{center}
    \label{table:CNR}
\end{table}

\section{Discussion}

\begin{figure}
    \centering
    \includegraphics[width=0.48\textwidth]{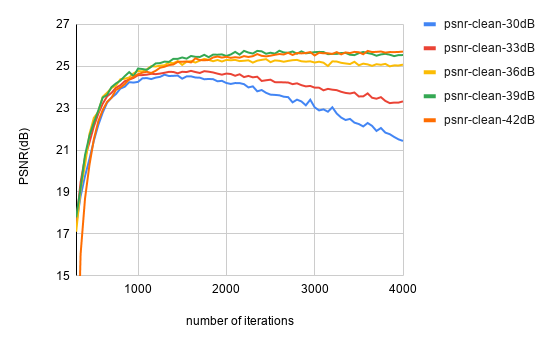}
    \includegraphics[width=0.45\textwidth]{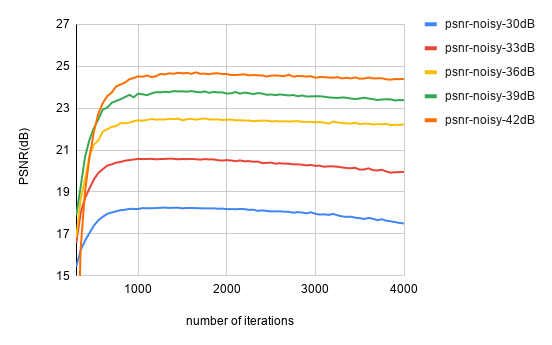}
    \caption{ At the graph above, the PSNR values of the reconstructions with the ground truth data and at the graph below, the PSNR values of the reconstructions with the noisy initial estimations are given.}
    \label{fig:noise-level-graph}
\end{figure}

The noise level on the projections is one of the parameters which affect the optimization curves and performance of the proposed method. Therefore, experiments are also done for different noise levels. In Fig. \ref{fig:noise-level-graph}, PSNR curves during the optimization process for different noise levels with 64 projections of an ellipses image are given. As the number of iterations increases, the output of the neural network starts to converge the noise in the projections by generating noisy reconstructions. Therefore, early stopping criterion is required especially for low SNR values such as $30$ dB and $33$ dB. The experiments investigating the effect of noise level in the reconstruction performance of the network suggest that noise level is a crucial issue and the iteration number should be tuned according to it.

The network structure is empirically determined as SkipNet. However, the number of scales is one of the crucial parameters to tune the performance of the network. \color{black}To compare different options, the DNNs (\textit{v1, v2, v3}) with different scale numbers were created. The channel depths in \textit{v2} architecture were chosen as the 2nd to 5th scales of channel depths in \textit{v1} architecture and \textit{v3} architecture is similarly selected as the channel depths in the v2 architecture from the 2nd to the 4th scales. The channel depths of three different network architectures are defined as follows:

\begin{itemize}
    \item \textit{v1}: $5$ scales, $[16, 32, 64, 128, 256]$.
    \item \textit{v2}: $4$ scales, $[32, 64, 128, 256]$.
    \item \textit{v3}: $3$ scales, $[64, 128, 256]$.
\end{itemize}
\color{black}

\begin{figure}
    \centering
    \includegraphics[width=0.48\textwidth]{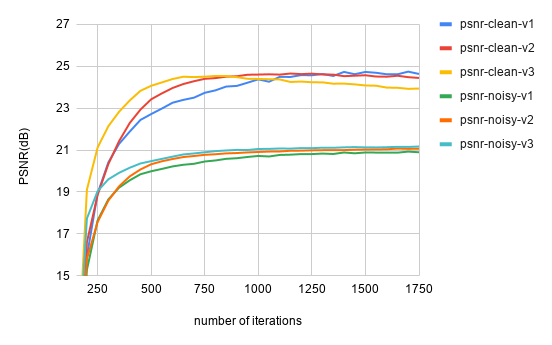}
    \includegraphics[width=0.48\textwidth]{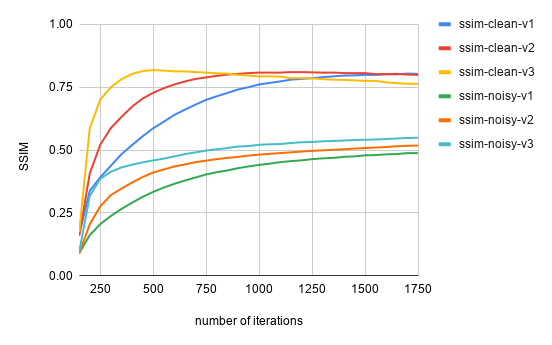}
    \caption{ PSNR and SSIM values for 3 different network architectures (\textit{v1}, \textit{v2}, \textit{v3}) with noisy initial reconstructions and ground truth (clean) images. }
    \label{fig:net-arc-graph}
\end{figure}

In Fig. \ref{fig:net-arc-graph}, the optimization curves for PSNR and SSIM values of the architectures (\textit{v1}, \textit{v2}, \textit{v3}) are given. The number of scales has a big effect on optimization curves in general. PSNR is maximized at different iteration numbers for each architecture. As the model gets deeper, overfitting starts at higher iterations numbers. \textit{v3} does not overfit even after $1750$ iterations. SSIM curves also have similar patterns with PSNR curves except for the differences between the architectures are more significant. The proposed DGR method is an unsupervised method and the layer depth of the network is one of the most crucial hyperparameters for the reconstruction performance and it should be tuned according to the noise level on the projections.

\color{black}The proposed methods have four main caveats.\color{black}

\begin{itemize}
    \item Reconstruction Speed: Because of the working principle of the proposed method, a deep neural network is trained during the reconstruction procedure. It can take up to $4000$ iterations which takes much more time from its alternatives such as iterative methods or inference of trained DNN. 
    \item Completely Unsupervised: The method does not take advantage of any prior data from the same domain. Although it is considerably hard to collect data for medical images to train a deep neural network, the proposed method does not exploit the existing limited data. It can be the topic of future studies in this area.
    \item \color{black} Selection of Weights: DNN based methods in common has a drawback of hyperparameter dependency. In our study, there are three different losses with different weights. Considering the precision of each weight as $0.001$, the span of parameter space is in the order of $10^9$ which makes it difficult to find the optimum weights for the most ideal reconstruction performance. \color{black} 
    \item Reconstruction Stability: Since the deep generative networks are fairly complex structures, it is nearly impossible to estimate convergence of the learning process. It can be estimated by empirical experiments. Through our experiments, collapses during the optimizations are observed occasionally and the results of the proposed method are stochastic due to the randomness existing in several parts of the method such as random input noise, randomly initialized network parameters, and dropout layers. They can vary from experiments to experiments with relatively high variance. Therefore, to compare the performance of the methods fairly, the numeric results of the methods are tested on several images and they are given with average and standard deviation values.
\end{itemize}

Considering the clinical potential of the proposed method, it produces superior results than the traditional method FBP, a commonly used iterarive method SART and the state-of-the-art regularized iterative method SART+TV for both analytical and human CT data, which makes it a promising alternative to be used in the clinical appliacations. However, it has two limitations in this manner as it is sensitive to the weights of the loss function terms and its reconstruction takes considerably longer time than the alternatives. Reconstruction time of the proposed method is roughly around ten minutes while its alternatives take less than one minute. However, it can be significantly decreased with more powerful GPUs.

\section{Conclusion}

We proposed an unsupervised DL reconstruction method for low-dose CT imaging. The proposed DGR method is noise redundant and not too much hyperparameter dependent for a certain noise level and number of view settings. Moreover, DGR method does not require any training on big datasets. Therefore, it is a favorable candidate for any domain in which collecting training images as labels is difficult. 

The performance of the proposed method is evaluated with analytical phantoms and human CT datasets by comparing to FBP, SART and SART+TV both qualitatively in visual comparison and 1D profiles senses and quantitatively in PSNR, SSIM, and CNR senses. Visual comparison and 1D profiles show that the proposed method provides the most acceptable results by preserving fine details with sharper edges and suppressing background noise. The proposed method with different loss function variants provides the best PSNR and SSIM performances for all view settings. 

One of the other criteria, for the evaluation of medical image reconstruction problems, is their clinical usability. To accomplish this, domain experts, in our case radiologists should also qualify the results. Radiologists' evaluations should be incorporated during the hyperparameter optimization of the method. For further studies, radiologists' opinions could be favorable to utilize our algorithm for its clinical usage.

\bibliographystyle{IEEEtran}
\bibliography{bibliography/ct,bibliography/ctdl,bibliography/dip,bibliography/dl,bibliography/metric,bibliography/n2n}

\end{document}